\documentclass[11pt,aas2pp4]{aastex}
\shorttitle{Cold dust structures in \CenA}
\shortauthors{Leeuw et al.}
\newcommand{\Th}{\mbox{This paper}}
\newcommand{\sTh}{\mbox{this paper}}
\newcommand{\um}{\,$\mu$m}

\newcommand{\CenA}{\mbox{Centaurus A}}
\begin{document}

\title{Deep Submillimeter Imaging of Dust Structures in \CenA} 

\author{Lerothodi
L. Leeuw\altaffilmark{1,2,3}
 Tim G. Hawarden\altaffilmark{3}, Henry
E. Matthews\altaffilmark{3,4}, \\ E. Ian
Robson\altaffilmark{3} and Andreas Eckart\altaffilmark{5}} 


\altaffiltext{1}{Centre for Astrophysics, University of Central Lancashire,
Preston PR1 2HE, England}

\altaffiltext{2}{Email: lleeuw@uclan.ac.uk}

\altaffiltext{3}{Joint Astronomy Centre, 660 N. A'oh\={o}k\={u} Place,
Hilo, HI 96720}

\altaffiltext{4}{National Research Council of Canada, 
Herzberg Institute of Astrophysics, \\ 5071 West Saanich Road, 
Victoria, BC, V9E 2E7, Canada}

\altaffiltext{5}{ Universit\"at zu K\"oln, I. Physikalisches Institut,
 Z\"ulpicherstr. 77, 50937 K\"oln, Germany}


\begin{abstract}
Images covering the central $450\times100''$ ($\sim
8.0\times2.0$\,kpc) of NGC\,5128 (\CenA) obtained using SCUBA at 850 and
450\micron\ with beam sizes of 14.5 and 8\arcsec\ respectively, are
presented.  These data are
compared with those obtained at other wavelengths, in
particular the optical, mid-infrared, and far-infrared continuum.
The sensitive 850 and 450\micron\ images show that the submillimeter (submm) 
continuum morphology and spectral index distribution of \CenA\
comprise four regions: an unresolved AGN core, an inner jet
interacting with gas in the dust lane, an inner disk of
radii $\sim 90''$, and colder outer dust.  The inner disk has a
high surface brightness, 
reverse-S-shaped feature in the 850 and 450\micron\ images that
coincides with the regions of intense 7 and 15\um\
continuum and a region of active star-formation.  The infrared (IR)
and submm images appear to reveal the same material as predicted by a
geometric warped disk model consisting of tilted rings.  We suggest
this scenario is more plausible than that recently proposed in literature
suggesting that the mid-IR emission in \CenA\ is primarily from a bar,
with a structure that is different from the extended warped disk
alone.  A dust mass total of $2.2
\times 10^6{\rm M}_{\odot}$ has been calculated within a radius of $
225''$, 45\% of which is in the
star-forming region of radius $\sim 90''$ about the nucleus. 

\end{abstract}


\section{Introduction}
\label{sec:intro}

NGC\,5128 (\CenA), at an assumed distance of $\sim 3.5$\,Mpc
\citep{hui93}, is the nearest giant elliptical galaxy and is
remarkable in several respects.  It is a
powerful radio source with a double-lobed structure extending approximately
$3^{\circ}.5 \times 8^{\circ}.5$ on the sky \citep{bol49,cla92,tin98}.
At the other extreme of scales a central source, less than 0.4 milliarcsec
($\sim$ 0.008\,pc) in extent \citep{kel97}, is prominent on radio through
X-ray images. This compact object feeds subparsec-scale relativistic outflows that are approximately aligned with, and clearly the generators
of, the vast outer radio lobes \citep[e.g.][]{tin98}.

The optical appearance is dominated by a dramatic warped dust lane at
least 12$'$.5 in E-W extent, which effectively bisects the main
body of the elliptical galaxy and almost completely obscures the nucleus
and all optical structure in the inner 500 parsecs \citep[e.g.][]{sch96}.
The outer isophotes of \CenA\ are markedly elongated in P.A. $\sim$
25$^{\circ}$. Faint shells, associated with both H\,\small{I} and CO emission, are
evident in these outlying parts \citep{mal83,sch94,cha00}. The somewhat
chaotic dust lane, and especially the shells, are strong evidence of a
relatively recent merger \citep{baa54,gra79, tub80, sch96, isr98} which is
generally believed (c.f. \citet{mar00} and references therein) to be
responsible for the nuclear activity.  H$\alpha$ and molecular line observations \citep[e.g.][]{van90} indicate
that the nucleus is surrounded by a rapidly rotating massive inner disk of
radius $\sim$2\,kpc with a pronounced warp \citep{nic92,qui92}, a scenario that is supported
by modelling of the structure of the obscuring dust seen in near-infrared
\citep{qui93} images.

Submillimetre wavelength emission, apparently thermal in origin, was first
observed from \CenA\ by \citet{cun86}, using the single-pixel bolometer
UKT14 mounted on the 3.8-m UK Infrared Telescope with a beam size of
$\sim 80\arcsec$. Millimeter-wave continuum observations by \citet{eck90}
in 22\arcsec\ and 45\arcsec\ beams show an unresolved nuclear source,
surrounded by extended thermal emission seen by $IRAS$ at 50 and
100\micron\, and which is roughly co-extensive with a region of strong CO
molecular line emission. More extensive submm continuum observations were
made by \citet{haw93}, who mapped the galaxy using the smaller
beamsize of $\sim 15\arcsec$ afforded by UKT14 mounted on the 15-m James Clerk Maxwell
Telescope (JCMT).\footnote{The James Clerk Maxwell Telescope is operated
by The Joint Astronomy Centre on behalf of the Particle Physics and
Astronomy Research Council of the United Kingdom, the Netherlands
Organization for Scientific Research, and the National Research Council of
Canada.} Although extended emission was visible in their images, these
were not sensitive enough to reveal much detailed structure beyond a
general extension corresponding to the optical dust lane and a brighter,
elongated, central feature, with an apparently thermal spectrum,
surrounding the strong, flat-spectrum nuclear source.

\Th\ presents new high-quality images of \CenA\ obtained at
submm wavelengths, that are sensitive in particular
to emission from cold interstellar dust.  These resolve the structure
of the inner disk  
down to scales of about $8\arcsec$ ($\sim 150$\,pc), while also
revealing details not 
previously seen of the faint outer dust emission at large radii.  
Recently, \citet{mir99} used ISOCAM, the infrared (IR) imager on the
{\it Infrared Space Observatory} ($ISO$),\footnote{$ISO$ 
is an ESA project with instruments funded by ESA member states (especially
the PI countries: France (for ISOCAM), Germany, the Netherlands and the
United Kingdom) with participation of ISAS and NASA.} to obtain images
at mid-IR 
wavelengths, and these results are compared, together with observations at
other wavelengths, with those presented in \sTh.  Re-processed
$IRAS$\footnote{The $IRAS$ data were re-processed using HIRES
routines at NASA/IPAC which is operated by the Jet Propulsion Laboratory,
Caltech, under contract to the National Aeronautics and Space
Administration.} data are also presented to complement the detailed
study of the dust properties in the disk.
 
\section{Observations}
\label{sec:obs}

Centaurus A presents special challenges for observations at submm
wavelength from Mauna Kea since it never rises more than 28\arcdeg\ above
the southern horizon. Images were nevertheless obtained using the
submm wavelength bolometer array SCUBA \citep{hol99} during several
observing periods with the JCMT during 1998.

The simultaneous 850 and 450\micron\ images were obtained during three
nights (UT 1998 March 29, April 9 and April 11) during a period in which
sky conditions were exceptionally transparent, with zenith opacities often
as low as 0.1 and 0.4 or better at 850 and 450\micron\, respectively. The
data were taken using the ``jiggle-mapping'' mode of SCUBA, which provides
a densely-oversampled image with a spacing of 2$\arcsec$.18, and each
observation results in a field about 2$\arcmin$.3 in diameter.  The array
orientation rotates with respect to the sky as the observation progresses,
further sampling the image plane.  A series of seven
overlapping field centers were observed, placed 55\arcsec\ apart along a
line at a position angle of 120\arcdeg\ (North through East), roughly
corresponding to that of the dust lane of \CenA. The chop throw was
120\arcsec\ perpendicular to this line.

The strong overlap (more than 50\%) between neighboring fields compensated
for the limited rotation of \CenA\ with respect to the
SCUBA arrays, offsetting the effect on the final reduced images of
occasional excessively noisy pixels.  A different subset of the seven
fields was observed on each of the three nights. Since the bright nuclear
source appears on the central three fields, it was observed at least twice
per night, which allowed corrections for pointing drifts to be measured
and applied to the final images.

To determine the beam pattern of the JCMT, the bright unresolved blazar
3C\,279 was observed each night using the same jiggle-pattern mode as
applied to \CenA.  Atmospheric opacities were determined from skydips made at intervals
during the observing, and instrumental gains were derived from images
of the JCMT secondary calibrators CRL\,618, IRC+10216 and
IRAS\,16293$-$2422 (Sandell 1994, 1998).  The imaging data analysis
was undertaken using the dedicated   
SCUBA data reduction software SURF \citep{jen98}, as well as KAPPA, GAIA
and CONVERT software packages provided by the Starlink Project\footnote{The Starlink Project is run by the
Council for the
Central Laboratory of the Research Councils on behalf
of the Particle Physics and Astronomy Research Council of the United
Kingdom.}.  The 
data reduction consisted of first flatfielding the array images, and 
then correcting for atmospheric extinction.  Next, pixels
significantly noisier than the mean were blanked-out; and, after initial inspection of raw images, pixels
containing relatively little flux from the source were used to correct
for correlated sky noise in each individual jiggle-map.  Corrections for pointing drifts were incorporated in the final images
using the fields in which the core source appeared. The apparent core
source size was reduced from 9.24 to 8\arcsec.42 (FWHM) at 450\micron\
after the inclusion of these corrections.

An image at 850\micron\ had been previously obtained under somewhat
worse conditions on UT 1998 February 14, 15 and 16. This image was
made using 
the ``scan-mapping'' mode and covers an area 4\arcmin\ square centered on
the bright core source. Despite the lower quality of this dataset, it provides
a useful comparison dataset for the reality of features in the two
850\micron\ images.  The sky transmission was not good enough on this
occasion for the simultaneous 450\micron\ data to be useful.
The instrumental gains for the ``scan-mapping'' mode
observations are 20\% less than those typical for the
``jiggle-mapping'' mode; therefore, the gains have to be determined
separately for the individual modes.   The beam in the 850\micron\
``scan-mapping'' image of \CenA\ in this
paper is slightly larger than in the
``jiggle-mapping'' images and the photometry differences between
the two final maps are within and proportional to the uncertainties:
the percentage differences are larger in the less sensitive parts of
the maps. 

\section{Results and Discussion}
\label{sec:results_etc}

\subsection{The Submm Images and Comparisons with other
Wavelengths}
\label{sec:images_comp}

The images of \CenA\ at 850 and 450\micron\ obtained from the combined
``jiggle-map'' datasets are about 6\arcmin\ long and 2\arcmin\ wide
(respectively 7\,kpc and 1.7\,kpc at \CenA) and cover much of the dust
disk. The 850 and 450\micron\ maps, which respectively have rms noise
values of 5 and 20\,mJy/beam, are shown in Figure~\ref{fig:jiggles}. 
Figure~\ref{fig:scanmap} shows the 850\um\ image obtained
from the ``scan-mapping'' observations, which has an rms noise 
of 25\,mJy/beam. 

The submm images show extended emission centered on the nucleus of \CenA\
and oriented at a position angle of about 115\arcdeg\ (North through East)
on the plane of the sky, in roughly the same orientation as the
prominent optical dust lane.  Within about $90''$ at both 850 and
450\micron, the emission
has a much higher surface brightness than at larger radii.  The bright
structures have a reversed S-shape suggestive of spiral structure or, as
proposed by \citet{mir99}, a bar.

The central source has a flux density per beam at least
40 times brighter than the surrounding emission from the dust lane, as
shown in the profiles along the major axis in Figure~\ref{fig:profiles}.    
Furthermore, the close resemblance of the central source
to the profiles of the JCMT beam (line with crosses in
Figure~\ref{fig:profiles}), shows the source is unresolved.
The point-source core has been isolated from the extended emission and
separate submm-to-IR spectral energy distributions 
(hereafter SED) have been constructed for the core and extended
galactic emission.  These are discussed in Section~\ref{sec:disk}.

Figure~\ref{fig:overlays} shows the 450\micron\ contours, our highest
resolution new SCUBA image, superimposed on a) an optical waveband (395
and 540\,nm) image
 courtesy of the
Anglo-Australian Observatory\footnote{Original plate material for the
Digitised Sky Survey in the southern sky
is copyright (c) the Anglo-Australian Observatory and was used, with their
permission, to produce the The Digitised Sky Survey at the Space Telescope
Science Institute under US Government Grant NAG W-2166.} and b)
the ISOCAM 7\micron\ image.  We now discuss these
comparisons, as well as what we learn from the 450/850\micron\
spectral index distribution map, in more detail. 

\subsubsection{Submm vs. Optical Morphology}
\label{sec:vs_optical}

The well-defined reverse-S-shape structure seen in high
surface brightness submm emission, within 90\arcsec\ of the galaxy
nucleus, is largely indiscernible in optical maps, implying that the
denser material around the nucleus, seen in the submm-wavelength images
is heavily
obscured at optical wavelengths.  However, the southern edge of the SE
high surface brightness submm emission aligns with the
southern ridge of the optical dust lane, in turn suggesting that the SE 
reverse-S-shape structure is on the near-side and therefore relatively less 
obscured in the optical, as reported by \citet{blo00} and
\citet{qui93} and based on their analyses of
mid-infrared versus $V$ and near-infrared data respectively.
Furthermore, if the submm/optical morphology is a 
manifestation of a warped dusty disk or spiral structure that is
highly inclined to the plane of the
sky (c.f. Section~\ref{sec:image_impl}), the southern edge is
consistent with an inner fold of the disk seen tangent to the 
line of sight.  Such a ridge would have a high column density in
the line of sight, as evident in the aligned southern edges of {\em high}
surface brightness submm emission and the {\em dark} optical lane.

A very rough
estimate can be made of the optical depths detected in the SCUBA
images, allowing a quantitative comparison
between the submm emission and
optical obscuration.  Using the emissivity (i.e. grain absorption coefficient)
from Hildebrand (1983) and assuming a grain temperature of 17\,K
(c.f. the diffuse dust in the Galaxy), a $V$-band optical depth of about 10
can be inferred for a surface brightness of $\sim 30$\,mJy/beam at
850\micron, as seen in the low-level emission detected in the current
submm images.  Dust clouds that
are evident in the optical image presented in \sTh\ are
expected to have $V$-band
optical depths of $\sim 10$.  Therefore, under reasonable assumptions,
the dust lanes seen in the optical obscuration map exhibited here
should be detected in the SCUBA images. 

Figure~\ref{fig:overlays} shows that the low surface brightness submm
emission generally follows the optical dust
absorption distribution, including the clockwise twists in the east
and west of the dust lane.  This submm emission corresponds
especially well with the dark
optical obscuration in the NW and northern part of the SE dust lanes,
and also with the very dark twist in the east, 
indicating that some low level submm emission must arise from the optically 
prominent dust material.  There is no marked difference between the
flux densities of submm low level emission that corresponds with the dark
optical obscuration in the northern edge of the SE dust lane and the
low level emission in the
southern ridge of the dust lane: in both regions, the 
flux densities are in the range $\sim 40$ to $\sim
120$\,mJy/beam and $\sim 10$ to $\sim 60$\,mJy/beam at 
450\,\micron\ and 850\,\micron\ respectively.  
\citet{sch96} find that the extinction along the 
southern ridge is much more than that in the northern edge of the SE
dust lane, where it reduces the $R$
flux by a factor of 6 compared to a factor of only 1.5 to 2 in the
northern edge of the SE lane.  
Therefore, the similarity of the flux densities in the northern and
southern low level submm 
emission despite the higher extinction in the southern ridge indicates
that the northern edge of the SE dust lane is well in the foreground,
in a region 
of low stellar density, and is thus heated by a more dilute stellar radiation 
field.

Within the length (major diameter) of $270\arcsec$ that excludes the
clockwise twists, low surface brightness submm emission
extends farther north and south beyond the optical feature: 
the average width (minor diameter) of the dust feature seen in submm
emission is $>\sim
90\arcsec$ compared with the optical obscuration of $\sim < 60\arcsec$
(see Figure~\ref{fig:overlays}(a)).  
The extension is appreciably marked in the south, where the dust giving
rise to the submm emission here
lies on the far-side of the galaxy and is overlain by the stellar
body.  Consistent with this scenario, when \citet{sch96} removed the
effects of foreground dust obscuration from $HST$ $I'$ band images, they
found a band of residual obscuration, presumably caused by dust
within the stars on the farther side of the galaxy.  

While it is clear
that the maximum width at which the dust lane will be 
detected in extinction and seen in thermal emission will depend on the
relative sensitivity of the two techniques, the width at which the 
feature is detected in the submm observations ($> \sim  90\arcsec $,
c.f. Figure~\ref{fig:overlays}(a)) 
surpasses the maximum width detected in all optical observations we are 
currently aware of, including the recent $HST$ \citep{sch96, mar00} and VLT\footnote{VLT 
astronomical image gallery is available at
http://www.eso.org/outreach/info-events/ut1fl/astroimages.html}
observations ($  \sim < 60 \arcsec $), supporting the
assertion that the low-level submm emission extends well beyond the optical 
obscuration.

\subsubsection{Submm vs. $ISO$ Mid-IR Morphology}
\label{sec:vs_ISO}

\citet{mir99} presented images of \CenA\ at 7 and 15\micron\ obtained with
ISOCAM as well as early
SCUBA ``jiggle-mapping'' images of a 2\arcmin\ field around the core of
the galaxy at 450 and 850$\mu$m. They note
the similarity of their IR (ISOCAM) and submm (SCUBA) images (which are much
less sensitive to faint emission than those presented here), observing
that there was the ``same general distribution'' between the ``warm'' and
``cold'' dust seen respectively in the ISOCAM and the early SCUBA image
published by them. \citet{mir99} conclude that the absence of submm
emission from the optical dark lanes ``is not due to major differences
between the spatial distributions of the cold and very warm dust
components," attributing the optical dust features to ``small amounts of
cold dust in the outer parts" of the system.

The SCUBA data exhibited in \sTh, which have much greater sensitivity and extend much
farther from the nucleus than the images published by \citet{mir99}, show
that there are \emph{remarkable} similarities between the
appearance of \CenA\ in the mid-IR and the submm (see
Figure~\ref{fig:jiggles}, Figure~\ref{fig:overlays}(b)). This is 
particularly true of the 450 and 7\micron\ images, even
though their wavelengths differ by a factor of 65.  Both images show the
reverse-S-shaped, high surface brightness structure out to 90\arcsec\ from
the core and the fainter extensions of this structure out to 120\arcsec.
The major {\it difference} is that the submm emission is seen to much
larger angular distances than is the mid-IR, and as pointed out
above, faint submm emission from some of the more outlying dust in the
optical dust lane is evident.

The contours in Figure~\ref{fig:overlays}(b) show a feature extending to about
15\arcsec\ radius from the nucleus in position angle 145\arcdeg; this may
correspond to that attributed to a circumnuclear torus by
\citet{haw93,isr98} and \citet{bryxx}.  The SCUBA and ISOCAM
images suggest it may represent
the inner folds of the reverse-S-shaped structure (or warped disk,
c.f. Section~\ref{sec:image_impl}) mentioned above.

\subsubsection{The Spectral Index Distribution}
\label{sec:SED}

Using the 450 and 850\micron\ imaging data, the
global spectral index distribution of \CenA\ is derived at
submm wavelengths and
used to delineate the submm components in the nuclear regions of
the galaxy as well as to investigate the dust properties of the
extended emission.  The following procedure was used.  The 450\micron\ data were first smoothed 
to the $14''$ resolution of the 850\micron\ map.  The spectral index
$\alpha$, where $S_{\nu} \propto {\nu}^{\alpha}$, between 450\micron\
and 850\micron\ was then computed as
\begin{equation}
\alpha
=\log{\left[\frac{S_{450}}{S_{850}}\right]}\bigg{/}\log{\left[\frac{850}{450}\right]}, \label{eqn:alpha}
\end{equation}
\noindent
where $S$ is the flux density per beam at each point in the map.  The
uncertainty in the spectral index is then given by
\begin{equation}
\left(\Delta\alpha\right)^2 = \left[{
                 \left(\frac{\Delta S_{450}}{S_{450}}\right)^2 + 
                 \left(\frac{\Delta S_{850}}{S_{850}}\right)^2}\right]
                 \Bigg{/}
                \left(\log{\frac{850}{450}}\right)^2,
\label{eqn:dalpha}
\end{equation} 
\noindent
where the variables are as for equation~\ref{eqn:alpha}. The largest
probable uncertainty in $\Delta S$ arises from the calibration, especially
for the 450\micron\ flux densities.  The maximum uncertainty in
our final map is estimated as ${\mid \Delta \alpha \mid}_{\rm{max}}
\approx 1.5$
and the mean uncertainty as ${\mid \Delta \alpha \mid}_{\rm{mean}} \approx
0.5$.  The uncertainty in the regions with $\alpha > 3.2 $ is most probably
underestimated as this is also the region of the map with the least
sensitivity.

The spectral index map of \CenA\ between 450 and 850\micron\
is shown in Figure~\ref{fig:spec}.  There are four regions apparent in
the map.  Contours at the spectral indices 2.0 (dashed lines) and 3.0
and 3.3 (solid lines) are overlayed to guide the eye to the less
obviously manifested 
features.  In the nuclear area, the unresolved core and a
feature apparently representing the inner jet are
distinguished.  Farther from the nucleus we
can discern, with progressively increasing spectral indices, the bright,
reverse-S-shaped structure, familiar from the intensity maps,
and the fainter outer dust, roughly corresponding to the optical dust lane
(see Section~\ref{sec:vs_optical}).

The spectral index of the \textbf{unresolved core} is close to zero,
consistent with the nonthermal spectrum from an AGN nucleus.  This
paper confirms
that the core source has a flat submm spectrum (i.e. $\alpha \approx 0$),
which is common for BL Lacs (blazars) \citep[e.g.][]{bro89}, and as
\citet{haw93} and \citet{kel97} have noted, suggests that \CenA\ may harbor a
low-luminosity blazar.

The feature which \sTh\ refers to as the \textbf{inner jet} is not evident on
the intensity maps but is prominent in Figure~\ref{fig:spec} as
(green-blue) areas of low spectral index to the NE and SW of the nucleus. The
spectral index of these features is in the range $1.5 \leq \alpha \leq
2.5$ (see dashed contour lines).  The northern extension is coincident with the inner radio jet
\citep{cla92}, which, as implied by the designation in \sTh, is not considered to
be merely by chance. However, the observed spectral index is much
higher than would
be expected from extrapolation of the power-law spectrum of the radio jet
itself.  One possible explanation is that the submm spectrum is not a
manifestation of the electron distribution of the relativistic
radio jet itself (or the jet overlaid by a warm dust component), but
rather of some internal mechanism (e.g. ionization of gas) resulting
from the interaction of the jet and the ISM in the inner regions of the galaxy.

\citet{bro83} and others have proposed that optical jets in galaxies such
as \CenA\ may be due to emission from interstellar gas that has been
entrained and heated by the flow of relativistic particles from the
nucleus.  The entrained gas would generate free-free continuum emission
with spectral index, in the optically thick case, of $\alpha \approx
2$, consistent with the observed spectral range of the
inner jet $1.5 \leq \alpha \leq 2.5$. We attribute the southern nuclear
extension in our submm spectral map to the same mechanism, though in this
case associated with the fainter counter jet recently reported by
\citet{tin98} and \citet{who00}, respectively from their high resolution
VLBI radio and $Chandra$ X-ray images.

A similar interaction of the jet and interstellar medium was
proposed by \citet{joy91} based on their near-IR observation of the
inner jet of \CenA, which manifested itself as a region of bluer color,
with spectral index $1.3 \leq \alpha \leq 2.9$, coincident with the radio
jet. \citet{mar00} also reported a
`blue channel' coincident with the radio jet, but they attributed it to
low extinction resulting from relatively low concentration of dust that
has been mechanically `evacuated' by the jet. \Th\ disagrees with this
interpretation as there is no sign of a deficiency of emission on
the 450\micron\ map at this location: the inner jet feature of
Figure~\ref{fig:spec} arises from a relative enhancement of
emission at 850\micron\ in these locations. 

The spectral index of the {\bf inner disk} is in the range $2.5 \leq
\alpha \leq 3.2$, consistent with thermal emission from dust. The lowest
indices occur in an almost circular region about 30\arcsec\ in radius
around the
nucleus; this is coincident
with an area of hot molecular gas reported by
\citet{sch98}\footnote{STScI-PRC98-14 Electronic Press Release is
available at
http://oposite.stsci.edu/pubinfo/pr/1998/14/pr-photos.html} from their 
$HST$ observations.  Emission with a steeper spectrum ($2.6 \leq
\alpha \leq 3.2$, see solid contour lines) approximately traces the inner disk of brighter
structures seen in the 7 and 450\micron\ images (see
Section~\ref{sec:vs_ISO} and Figure~\ref{fig:overlays}b) out to $\sim$
90$\arcsec$.  Farther still from the nucleus, in the fainter {\bf outer
disk} the spectral index increases to $\alpha \sim 4$.  As underscored by the
solid line contours in Figure~\ref{fig:spec}, the spectral index is
asymmetric about the nucleus in the inner disk region of the
map: it is somewhat
flatter on the eastern part of the inner disk, indicating
slightly different relative
distributions of dust grains on either side of the nucleus.  

If the dust seen in the maps presented in \sTh\ is emitting with
spectral index $\alpha = 2 + \beta$ (where $\beta$ is the emissivity
index), the
spectral index in the extended region implies that for a given
temperature, the dust in \CenA\ has on average a low $\beta$., i.e. the
dust is made of relatively large grains, and that its temperature is
fairly cool.  Furthermore, the somewhat flatter spectral index in the
south-east suggests the dust in this region is made of smaller grains
with moderately warmer temperatures than those in the same region in
the NW.  This is not surprising as the SE is reported to
exhibit stronger H\,\small{I} emission \citep{van90}, weaker
$^{12}$CO(1-0) and stronger H\,$\alpha$ line emission compared to the
NW, that led \citet{eck90} to speculate that star-formation and its
associated high radiation field were greater in the SE.

The uncertainty in the emissivity analysis in our spectral index map (Figure 
~\ref{fig:spec}) does not permit us to quantitatively examine the
effects, if any, due to  diffuse ``cirrus'' grains in the map.  The
dust in the outer disk, where the
emissivity index $\beta > 1.5$, may well consist of ``cirrus'' grains, which, 
as it is generally accepted \citep[e.g.][]{row92}, would have an
emissivity index close to
$\beta=2$  and temperatures $15<T<40$\,K.
\citet{eck90} associated the far-IR emission outside the molecular
star-forming disk (ie. radius $ > \sim 90\,\arcsec $)
with ``cirrus'' clouds with scale height larger than that of the 
molecular gas disk.  They predicted that the submm
extended emission would also originate from dust constituting
``cirrus'' grains as could well be the case.  The large western and eastern
warps seen in the 850$\mu$m and 450$\mu$m images (see
Figure~\ref{fig:jiggles}) resemble the distribution of the \ion{H}{1}
emission \citep{sch94}, indicating that, if the
dust in the warp features (and the general outer disk) in \CenA\ is composed of
``cirrus'' grains,  the \ion{H}{1} emission traces the ``cirrus'' dust in
NGC\,5128 just as in the Milky Way. 

\subsection{Implications of the Multiwavelength Images}
\label{sec:image_impl}

Our submm images demonstrate that outlying dust in \CenA\ is significantly
cooler than the material in the bright elongated features within
$\sim 90\arcsec$ of the nucleus. This is naturally to be expected: at
larger radii the ambient radiation field heating the dust is more dilute
because of the much lower density of stars \citep{eck90,sch96,mar00}. The
emission from this dust generally follows the optical dust lane.

The warped disk model proposed by \citep{qui92} and explored by
\citet{qui93}, consisting of tilted rings of material, predicts the
structure of such warmer and colder material, respectively seen in the
mid-IR and far-IR to submm, rather well. This scenario is
supported by \citet{eck99}, who showed that a warped structure of tilted
rings explained not only the line emission in the disk of \CenA\
\citep{qui93}, but also a complex absorption-line system towards the
nucleus, without a need for any additional structures. 

From their $ISO$ 7 and 15\micron\ images, \citet{mir99} conclude from a
comparison with similar mid-IR images of the dwarf barred
spiral NGC\,1530 that the structure imaged in the mid-IR in \CenA\ is itself
a barred spiral: the prominent emission peaks $\sim 75\arcsec$ from the
nucleus at each end of the bright mid-IR and submm structures are
hypothesised to be foreshortened arms twisting anti-clockwise from the
outer ends of the bar. \citet{blo00} support this proposal from their
$VHK$ and 15\micron\ study of \CenA, noting too that the warm dust seen
by ISOCAM contributes little to the extinction seen at optical
wavelengths.

Kinematics is clearly important in understanding the true nature of the
near-nuclear structures in \CenA. \citet{mir99} address these in their
Figure 4, a comparison of the IR images with the CO kinematics in the form
of a position-velocity (P-V) plot from \citet{qui92}. On this figure they
identify the strong CO feature extending from $-1^{\arcmin}.2$,
300\,km\,s$^{-1}$ to $+1^{\arcmin}.2$, 800\,km\,s$^{-1}$ as the bar,
undergoing solid
body rotation, and weaker outlying features at nearly constant velocity at
larger radii as the ``typical'' rotation curve of galactic disks.

However, an alternative interpretation of the PV diagram is that
the ``bar'' represents a nearly-complete, highly foreshortened, ring of
material at a radius of $\sim 65\arcsec$ ($\sim 1300$\,pc).  In that
scenario the
concentrations of IR and submm emission NW and SE of the nucleus represent
the ends of the ring where the optical depth is maximized. Similarly, the
``EW high-velocity feature" mentioned by \citet{mir99}, which they note
does not fit their scenario, is readily explained by another ring of
molecular material at a radius of $\sim 20\arcsec$ ($\sim 400$\,pc), tilted
relative to the outer ring and evident on the mid-IR images as two small
extensions to the nuclear feature in PA 80$\arcdeg$ and 260$\arcdeg$.
Furthermore, the outer features of the PV diagram also have an alternative
interpretation: the feature $>$1$^{\arcmin}.7$SE of the nucleus has the
properties of an arc of material starting at radius of $\sim 4.6$\,kpc and
ending perhaps 2\,kpc from the nucleus close to the minor axis: in an
normal edge-on spiral, such a feature would be interpreted as a
spiral arm.

\citet{mar00} also accept the bar model since its edges appear to be
delineated by linear concentrations of star-forming regions seen in their
Pa\,$\alpha$ images, suggesting that these delineate the shocks
normally seen as dust lanes along the leading edges or the midlines of
galaxy bars in optical images.  However, they also noted that their data
do not rule out a warped ring without a bar, as the star-formation is
found in regions of the warp that are tangent to the line of sight, and if
the young stars are above or inside the disk, they would indeed appear to
be concentrated where the Pa\,$\alpha$ emission is seen.

We suggest the latter interpretation is more plausible for the following two
reasons.  First,
the same elongated distribution of Pa\,$\alpha$ emission regions, which
\citet{mar00} show to coincide precisely with the elongated mid-IR ridge
that passes just south of the nucleus, is equally coincident
with the dense, narrow, optical dust lane that passes south of the
nucleus in Figure~\ref{fig:overlays}(a). Since this is the {\em only}
feature of the bright central complex (apart from the nucleus itself)  
that is potentially identifiable with an optically-visible feature, we
suggest that it is not in fact part of the near-nuclear complex at all,
but, as suggested by the models of \citet{qui92}, a manifestation of an
outlying fold or ring in the warped disk.

Second, we note that the bar-shocks traced by dust in early-type field
spirals are asymmetric about their nuclei, since they lie in the leading
edges of the bars, while those in late type systems are centered in the
bars. However, the Pa\,$\alpha$ emission regions in Figure 13 of \citet{mar00}
pass by several arcsecs south of the nucleus from the SE to west of the
nucleus. This is not what would be expected of a shock in a bar in a field
spiral, but more the distribution to be expected if the star-forming
regions are associated with the trailing dust lanes in a typical spiral arm,
or in this case the dusty rings postulated by \citet{qui92}.

Therefore, the combined evidence to date from the CO
kinematics, the mid-IR images, the bright and faint submm features and the
Pa\,$\alpha$ images supports the warped rings model of
\citet{qui92} rather than indicating the presence of a true bar, and may
even represent spiral structure in the dust lane of \CenA.

One of the attractions of the bar scenario for both
\citet{mir99} and \citet{mar00} is its expected utility as a mover of
material from larger to smaller radii in the disk in order to fuel the
AGN.  However, since
the association of barred structure with the presence of an active
nucleus in field galaxies, though long sought, is marginal at best
(c.f., for example, \citet{ho96} and references therein), the
presence of the AGN cannot be taken as an argument for the reality of the
putative bar.  Not having a bar to fuel the active nucleus is not too
great a loss; e.g. recent work by \citet{dus00} suggests that viscosity in
a thin gas
plane may also provide an efficient AGN fuelling mechanism.   

\subsection{The Extended Emission Temperature and Dust Mass Estimates}
\label{sec:disk}

 \Th\ has shown that the central source in \CenA\ is unresolved and has a flux
density per beam at least 40 times brighter and a spectrum markedly flatter
than the surrounding emission from the dust lane; i.e. the central
source is clearly distinct from the extended emission
(c.f. Sections~\ref{sec:results_etc} \& ~\ref{sec:SED} and
Figure~\ref{fig:profiles} \& Figure~\ref{fig:spec}).  Therefore, an
analysis of the 
SED for the extended submm to
infrared emission in
\sTh\ is preceded by the isolation of the point-source core 
from the extended emission 
and the determination of the respective flux 
density estimates for these separate components.
  Adjustments are made when dealing with the JCMT and
 $ISO$ data, both taken with small beams, and the $IRAS$ data, which were
obtained with significantly larger beams.  

The core flux densities from 850 to 7\micron\ are presented in
Table~\ref{tbl:fluxes}, together
with integrated flux densities determined for two regions of the extended
emission from the dust lane: (1) an ellipse of $60 \times 180$\arcsec\
{\it minus}
the core measurement and (2) an elliptical annulus of $120-60 \times
450-180$\arcsec.  The far-IR flux densities are archival $IRAS$ data for \CenA\ that we
re-analysed using the HIRES facility to 
extract the best possible spatial resolution and to ensure that we
have the same registration of the regions from which we determine
integrated flux densities.  The ellipse and annulus are denoted respectively as
the {\em inner disk} and {\em outer disk}, as the two regions have
different submm continuum morphology and spectral
index distributions (see Sections~\ref{sec:vs_ISO} \&
~\ref{sec:SED}).  The SED for the two extended regions are plotted in
Figure~\ref{fig:disk_spectrum}, and temperatures as well as dust
masses determined for the emission from these regions are discussed
below.  

The compact core is unresolved on the submm and ISO mid-IR images (FWHM
$<8''$ at 450\micron, $<4''$ at 7\micron), as would be expected if it
is predominantly a manifestation of the
unresolved central radio source \citep[c.f.][]{kel97}).
In the case of the $IRAS$ data, the best HIRES beams achieved in \sTh\ are still fairly large ($50''$ to
$90''$) so that the core extent is impossible to constrain and its flux density
not easily separable from that of the extended emission.  Therefore,
the ellipse $IRAS$ flux densities
listed in the table are simply the integrated
measurements minus the core flux density of $8 \pm 2$\,Jy, at both
100$\mu$m and 60$\mu$m, which is an extrapolation from
the submm core SED ($\alpha_{81\,{\rm GHz}}^{877\,{\rm GHz}} = - 0.01
$, S$_{\nu} \propto \nu^{\alpha} $) to 60$\mu$m (5000\,GHz).  A full
discussion of the SED of the submm core in \CenA, and its
implications, is outside the scope of this paper; for an extensive
analysis of this matter, see, among many others, \citet{bai86, haw93,
kel97, ale99}; and \citet{mar00}. 

The integrated submm to far-IR flux densities are fit
by a two-component, optically-thin modified blackbody of the form 
\begin{equation}
S_{\nu} =  (1-{\rm{exp}(-{\lambda_o}/{\lambda}})^{\beta}) 
[\Omega_{1} B_{\nu}(T_1) + \Omega_{2} B_{\nu}(T_2)],  \label{eqn:greybody}
\end{equation} 
where $S_{\nu}$ is the observed flux density at frequency $\nu$, $\Omega$ the
solid angle for 
the modified blackbody component, $B_{\nu}(T)$ the Planck function at
temperature $T$, $\lambda_o$ the wavelength at which the optical depth
is unity and $\beta$ the emissivity index of the grains.  First, a
12\,K and 40\,K two-component modified blackbody with
dust emissivity index of 1.3 fits the
data for the {\em inner disk}, and secondly, a
12\,K and 30\,K modified blackbody with dust emissivity index of 1.6 fits the
data for the {\em outer disk} (c.f. Figure~\ref{fig:disk_spectrum}).
These fits, in
particular the derived relative emissivities
and temperatures, are consistent with the spectral index maps, showing
that the average dust temperature decreases from the galactic core
outwards.  Unsurprisingly, these temperatures are also consistent with
the 42\,K and 32\,K determined by \citet{eck90} from their 50 and
100\micron\ $IRAS$ flux densities alone, as well as the average
$\sim 40$\,K determined by \citet{ung00} from their 40 to 100\micron\
{\it ISO} LWS maps of the {\it inner disk} region.

The mass of emitting dust $M_{\rm d}$ can then be estimated from
\citep[e.g.][]{hil83} 
\begin{equation}
M_{\rm d}= \frac{{S_{\nu} D^2}}{{k_{\rm d} B(\nu ,T)}},
\label{eqn:md}
\end{equation}
\noindent
where $S_{\nu}$ is the measured flux density at frequency $\nu$, $D$ is the
distance to the source, $B({\nu} ,T)$ the Planck function and $k_{\rm d} =
3Q_{\nu}/4a\rho$ the grain mass absorption coefficient where $a$ and
$\rho$ are respectively the grain radius and density. Values of
$k_{\rm d}^{450\,\mu {\rm m}} = 0.25\,{\rm m}^2 {\rm kg}^{-1}$
and $k_{\rm d}^{60\,\mu {\rm m}} = 3.3\,{\rm m}^2 {\rm kg}^{-1}$
\citep[e.g.][]{hil83} are assumed, yielding dust masses of $2.5 \times
10^4 {\rm M}_{\odot}$ for 
${\rm T}=40$\,K and $9.6 \times 10^5 {\rm M}_{\odot}$ for ${\rm
T}=12$\,K in the 
{\em inner disk} and dust masses of $4.9 \times 10^4 {\rm M}_{\odot}$ for ${\rm
T}=30$\,K and $1.2 \times 10^6 {\rm M}_{\odot}$ for ${\rm
T}=12$\,K in the {\em 
outer disk}. The total mass for the dust that emits from the far-IR
through the submm wavelength is then $9.9 \times 10^5 {\rm M}_{\odot}$ and $1.2
\times 10^6 {\rm M}_{\odot}$, respectively in the {\em inner disk} and {\em
outer disk}, giving an overall total of $2.2 \times 10^6{\rm
M}_{\odot}$ for the
dust lane region of the galaxy we have observed (ie. radii $< 225''$
or $\sim 4.5$\,kpc).

The advantage of using optically thin submm emission to 
determine the dust mass is, unfortunately, offset by the increased
uncertainty in the properties of dust and subsequently of $k_{\rm d}$,
as $\lambda$ is increased 
from the far-IR to submm wavelengths.  A
different choice of $k_{\rm d}$ could result in an estimation of
the dust mass that differs by a factor as large as $\sim 10$ \citep{dra90}. 
The dust mass derived in \sTh\ in a region of radius $\sim
90''$ (the {\em inner disk}) is at the
lower end of the dust mass of $1-2 \times 10^6 {\rm M}_{\odot}$
derived by \citet{blo00} from their $V-15\mu$m spatial distribution map  
in the same area.  This is surprising because they reported
the estimate was from dust in the ``bar+arms of a mini-spiral, excluding
any diffuse dust.''  However, given the uncertainty, the
mass in \sTh\ for the {\em  inner disk} dust is at least {\it
consistent} with \citet{blo00}. 

\section{Conclusions}

The SCUBA 850 and 450\micron\ images in \sTh\ show that the submm continuum
morphology and spectral index distribution of \CenA\ comprise:

\begin{enumerate} 

\item {\bf The Nucleus} and associated structures: A distinct, unresolved,
flat-spectrum AGN core with circumnuclear structure, including, to the
northeast and southwest, areas of low spectral index which we suggest
arise from free-free emission in ionised gas entrained in the nuclear
outflow;

\item {\bf Inner disk:} A prominent, elongated feature that may be a
circumnuclear ring (or bar) extends across the center of the galaxy out to
a radius of $\sim 90\arcsec$. It contains considerable real structure,
including concentrations of emission towards its ends and a
reverse-S-shaped twist; this structure appears reflected about the
nucleus. The details of this structure are very similar indeed to what is
seen on mid-IR $ISO$ images, so the cooler dust seen in the sub-mm is very
closely co-extensive with the much warmer material (transiently heated by
UV-photon absorption) seen at 7 and 15\,$\mu$m, and which establishes
that the {\it inner disk} is a locus of vigorous star formation. It is also a
strong source of CO emission with kinematics consistent with a ring or
rings of gas around the nucleus, rather than a bar, as has been
suggested elsewhere \citep{mir99}.

\item {\bf Outer disk and outlying dust features:} Low-level emission,
at least some of
which comes from the foreground material in the prominent optical dust
lane, extends to the eastern, and probably to the western, end of our
maps. It traces the clockwise twist of the optical feature. 

Some of the observed low-level emission does {\it not} coincide with
optically-visible dust. The sensitive SCUBA images are, for the
first time, showing the direct detection of emission arising from dust
that is in the far side of the galaxy, overlaid and obscured by the
stellar component of NGC\,5128, and thus not seen in optical images.

\end{enumerate}

A warped disk model consisting of tilted rings \citep{qui93} predicts the
structure of the warmer and colder material, especially in the {\em
inner disk}, rather well.  It appears that in this vicinity the IR
and submm images to a large extent reveal the same material; however, at
larger radii, the dust is cooler because it is immersed in a much less
intense stellar radiation field.  Alternative arguments are presented
in papers by 
\citet{mir99} and \citet{blo00}, who interpret the mid-IR (and brighter
submm) structures as a true barred spiral.

Using the continuum integrated flux densities from far-IR through submm
wavelengths, \sTh\ derives a total dust mass of $2.2 \times 10^6{\rm M}_{\odot}$
within a radius of $ 225''$ of the nucleus of \CenA.  About 45\% of the
dust mass is in the star-forming {\em inner disk} within about $< 90''$ of
the nucleus.

\section{Acknowledgements}

LLL thanks the University of Central Lancashire for a full-time research
studentship and the Joint Astronomy Centre for great hospitality while
working on this research.  We thank Felix Mirabel for the ISOCAM data
and acknowledge useful discussions with Alice Quillen, Wayne Holland,
John Whiteoak and Tim Cawthorne.  Tim Jenness 
and Malcolm Currie are also acknowledged for their advice in the full
exploitation of the SURF and other Starlink Project software packages
in our data analysis and displaying.  For some pointed and invaluable
comments that improved the final quality of this paper, the referee
and scientific editor are also acknowledged.


\figcaption[]{Images of \CenA\ obtained with SCUBA at 850\micron\
(top) and 450\micron\ (bottom), derived from a series of individual
``jiggle-mapping'' (see text) exposures offset along a line roughly
corresponding to the optical dust band of \CenA.  The 450\micron\ image
is smoothed slightly and the central source displayed saturated in
order to highlight the low-level, extended emission.
The panel key indices are in Jy/beam corresponding to color-coded
intensities, and the contour heights in the
850 and 450\micron\ images are
[0.01, 0.03, 0.05, 0.07, 0.1, 0.14, 0.2, 0.3, 0.4, 1.0, 2.5,  and 4.0\,Jy/beam] and
[0.04, 0.16, 0.28, 0.40, 0.52, 0.8,1.0, 1.5, 2.5,  and 3.5\,Jy/beam]
respectively.  \label{fig:jiggles}} 

\figcaption[]{Image of the central 4\arcmin\ at 850\micron\
obtained using with SCUBA and using the complementary ``scan-map''
technique.  The central source is displayed saturated in
order to highlight the low-level, extended emission.  This
map covers a smaller area than the ``jiggle-mapping'' set in
Figure~\ref{fig:jiggles}.  The panel key indices are in Jy/beam corresponding to color-coded
intensities, and the contour
heights are [0.07, 0.1, 0.2, 0.4, 1.0, 2.5,  and 4.0\,Jy/beam].  \label{fig:scanmap}}

\figcaption[]{Profiles of the emission from \CenA\ at 850 (top) and
450\micron\ (bottom) along an axis roughly coincident with the optical dust
lane.  The profile of the JCMT beam is superimposed in a line marked
with crosses. 
This beam profile is obtained from a map of the JCMT pointing source 3C\,279 
scaled to the peak flux of Centarus\,A and determined along an axis roughly 
coincident with the major axis of the dust lane.  \label{fig:profiles}}

\figcaption[]{Contours of the 450\micron\ image (see
Figure~\ref{fig:jiggles}) superposed on,
from top to bottom, (a) an optical waveband image of \CenA,
digitised from the $IIIaJ$ emulsion photographic plates which are sensitive
to emission between 395 and 540\,nm wavelengths,
courtesy of the Anglo-Australian Observatory, and (b) a 7\micron\
ISOCAM negative image \citep{mir99}.  The contour heights of the 450\micron\
emission in (a) and (b) are [0.04, 0.16, 0.28, 0.40, 0.52, 0.8, 1.0,
1.5, 2.5, and 3.5\,Jy/beam.] and [0.4, 0.7, 0.96, 1.2, 2.4, and
4.8\,Jy/beam], respectively.  In (a), the contours are of the slightly
smoothed image, presented as the 450\micron\ map in Figure~\ref{fig:jiggles}
to highlight the low surface brightness emission that corresponds with
the outlying parts of extended optical dust lane.  \label{fig:overlays}}  

\figcaption[]{Map of the spectral index distribution at submm
wavelengths, derived from the 450 and 850\micron\ data presented in
this paper (see Figure~\ref{fig:jiggles}).  Four regions are clearly
apparent in the map: the unresolved core (blue-black) and a
feature apparently representing the inner jet (green-blue) in the nuclear area; and
the bright, reverse-S-shaped {\em inner disk} (yellow-green-red), familiar from the intensity maps
(shown in Figure~\ref{fig:jiggles}), and the fainter {\em outer disk}
(red-yellow) farther from the nucleus.  Contour
heights at the spectral indices 2.0 (dashed lines) and 3.0 and 3.3  
(solid lines) are overlayed to guide the eye to the the inner jet and
inner disk features 
in the map.  The panel key indices correspond to color-coded
spectral indices obtained between 850 and 450\micron.  
\label{fig:spec}}
   
\figcaption[]{Integrated flux densities in two annuli centered on the core of
\CenA.  The error bars indicate the uncertainty in the flux density
measurements, which are dominated by the uncertainty in the flux
calibration and to a lesser extent the beam error-lobe contribution.
The top and bottom SEDs are respectively for the {\em
inner disk} of $60 \times 180$\arcsec\ minus the core flux density and the
{\em outer disk} annulus of $120-60 \times 450-180$\arcsec.  The submm to
far-IR data points are fit by two component optically thin
modified blackbodies of temperatures 12\,K and 40\,K with dust emissivity index
of 1.3 for the {\em inner disk} (top) and 12\,K and 30\,K with dust emissivity
index of 1.6 for the {\em outer disk} annulus (bottom). 
\label{fig:disk_spectrum}} 

\clearpage

\begin{table}
\begin{center}
\caption{Submm to Mid-IR Fluxes of the \CenA\ Core and Extended
Emission\label{tbl:fluxes}}
\begin{tabular}{lcccc}
\tableline\tableline
Filter & Frequency & Core\tablenotemark{a} & {\it Inner
disk}\tablenotemark{b} & {\it Outer disk}\tablenotemark{c} \\
	& (GHz) & (Jy) & (Jy) & (Jy) \\
\tableline
SCUBA\,850\tablenotemark{d}  & 350 &  8.1 $\pm$ 0.8 &   2.7 $\pm$ 0.4  &   3.4 $\pm$ 0.5 \\ 
SCUBA\,750\tablenotemark{d}  & 407 &  8.1 $\pm$ 1.6  & \nodata & \nodata \\
SCUBA\,450\tablenotemark{d}  & 667 &  7.9  $\pm$ 0.8 &  16.9 $\pm$ 3.2 &  22.6 $\pm$ 5.2\\ 
SCUBA\,350\tablenotemark{d}  & 866 &  7.7 $\pm$ 1.9  & \nodata & \nodata \\
$IRAS$\,100 & 3000 & \nodata & 119 $\pm$ 20\tablenotemark{e} & 181 $\pm$ 36 \\
$IRAS$\,60  & 5000 & \nodata &  96 $\pm$ 15\tablenotemark{e} &  77 $\pm$ 15 \\
$IRAS$\,25 & 12000 & \nodata &  15 $\pm$ 2\tablenotemark{f} &  6 $\pm$ 1 \\ 
ISOCAM\,15 & 20000 &  1.2 &  10.4 & \nodata  \\ 
$IRAS$\,12 & 25000 &  \nodata &   11 $\pm$ 2\tablenotemark{f} &   6 $\pm$ 1 \\
ISOCAM\,7 & 42857 &  0.7 &   9.4  & \nodata  \\ 
\tableline
\end{tabular}


\tablenotetext{a}{Peak fluxes of the unresolved core, which
in the case of the SCUBA and ISOCAM images was clearly distinct from
the extended emission.}
\tablenotetext{b}{The integrated fluxes determined in
an ellipse of $60 \times 180$\arcsec\ minus the core flux.}
\tablenotetext{c}{The integrated fluxes determined in
 an elliptical annulus of $120-60 \times 450-180$\arcsec.}
\tablenotetext{d}{The SCUBA filter bandwidths are all about 30\,GHz.}
\tablenotetext{e}{In the case of these 100 and 60\,$\mu$m
ellipse fluxes, the core fluxes subtracted off are an extrapolation of the
submm core 
flat SED ($\alpha_{81\,{\rm GHz}}^{877\,{\rm GHz}} = - 0.01 $, S$_{\nu} \propto
\nu^{\alpha} $) to 60\,$\mu$m (5000\,GHz), as the HIRES beams are very large
($50''$ to $90''$) so that it was impossible to isolate the core
measurements from the extended emission.} 
\tablenotetext{f}{Core fluxes have not been subtracted from these
integrated fluxes as no reasonable estimate of the core measurements at these
wavelengths could be extrapolated from our present data.}
\tablecomments{The $IRAS$ data were re-processed using HIRES
routines at NASA/IPAC.}

\end{center}
\end{table}

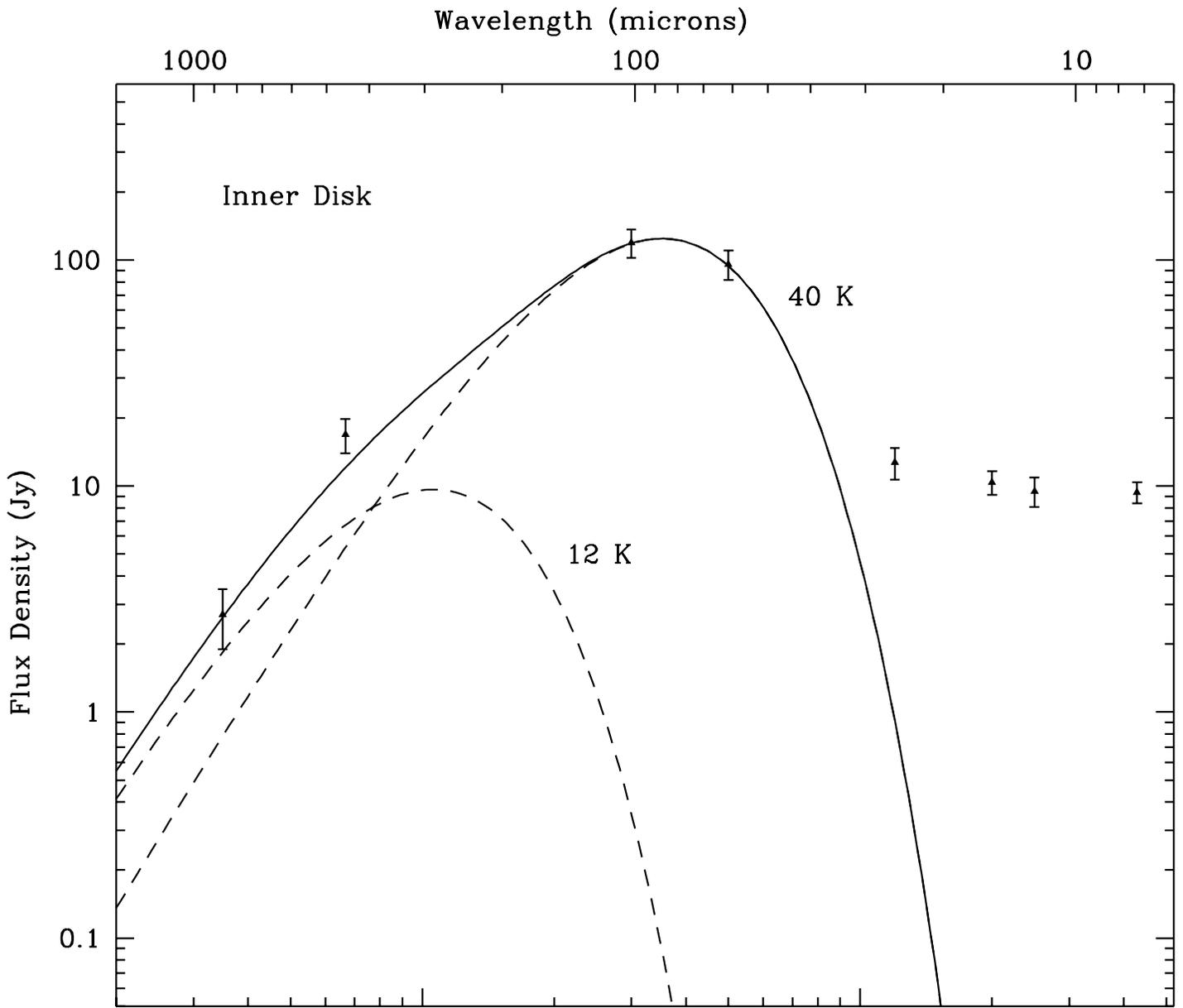
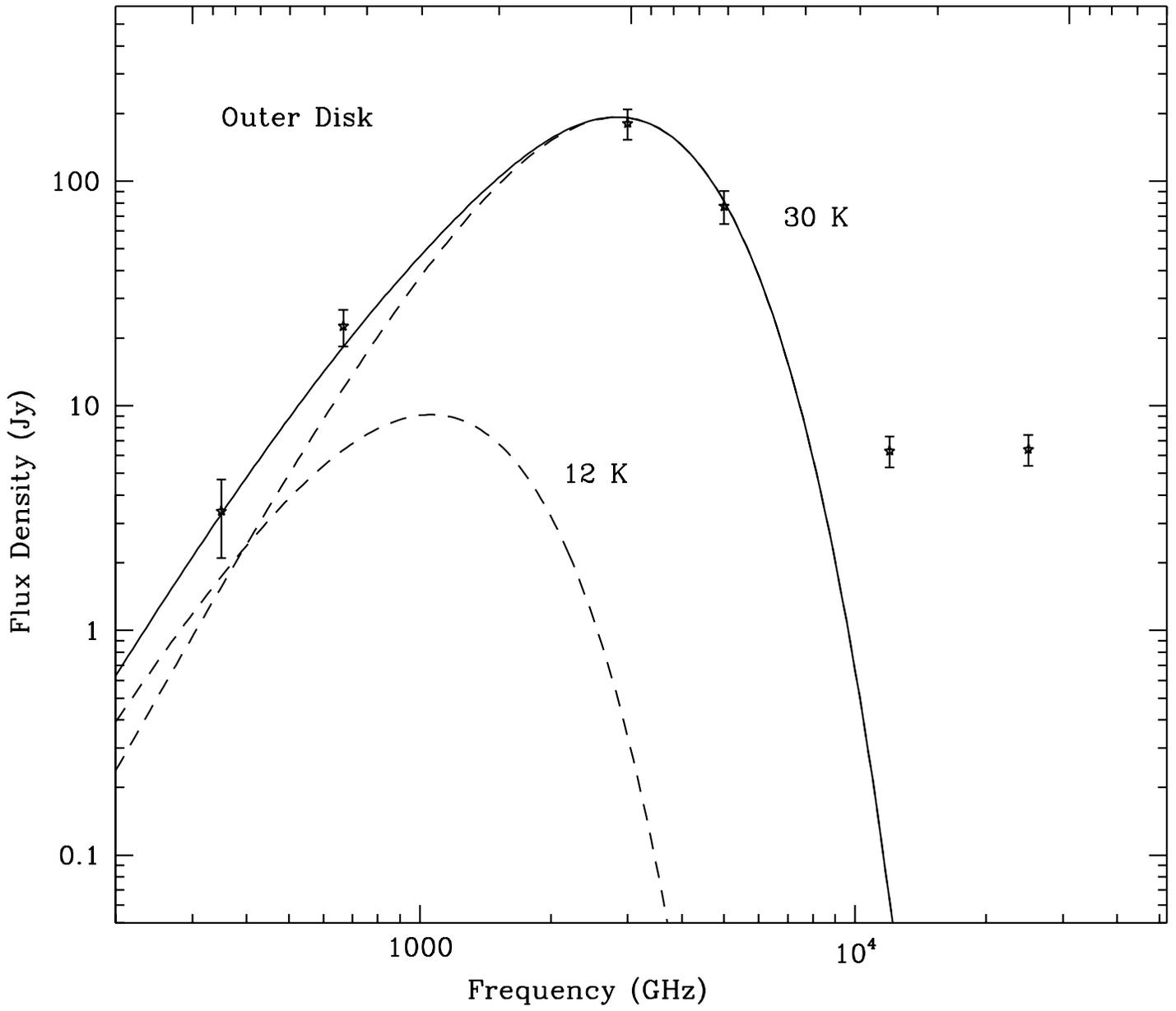
\end{document}